Photochromic response of encapsulated oxygen-containing yttrium hydride thin films

*Marcos V. Moro[1,*], Sigurbjörn M. Aðalsteinsson[1], Tuan. T. Tran[1], Dmitrii Moldarev[1, 2], Ayan Samanta[3], Max Wolff[1] and Daniel Primetzhofer[1]*

[1]Department of Physics and Astronomy, Uppsala University, Box 516, 751 20 Uppsala, Sweden
[2]Department of Materials Science, National Research Nuclear University, Kashirskoe shosse 31, Moscow, Russia.
[3]Department of Chemistry, Ångström Laboratory, Uppsala University, Box 538, 751 20 Uppsala, Sweden

[*]Corresponding author: marcos.moro@physics.uu.se

**Abstract**

Photochromic oxygen-containing yttrium-hydride thin films are synthesized by argon-magnetron sputtering on microscope slides. Some of them are encapsulated with a thin, transparent and non-photochromic diffusion-barrier layer of either $Al_2O_3$ or $Si_3N_4$. Ion beam-based methods prove that these protective diffusion barriers are stable and free from pinholes, with thicknesses of only a few tens of nanometers. Optical spectrophotometry reveals that the photochromic response and relaxation time for both – protected and unprotected – samples are almost identical. Ageing effects in the unprotected films lead to degradation of the photochromic performance (self-delamination) while the photochromic response for the encapsulated films is stable. Our results show that the environment does not play a decisive role for the photochromic process and encapsulation of oxygen containing rare-earth hydride films with transparent and non-organic thin diffusion barrier layers provides long-time stability of the films, mandatory for applications as photochromic coatings on e.g., smart windows.

## 1. Introduction

Recently, it was found that oxygen-containing rare-earth metal hydride thin films (REHO) change their optical transmission under illumination with visible light at ambient conditions [1–5], attracting the attention of the scientific community for their potential applications in next generation of photochromic coatings (smart-windows) and optoelectronic devices (gas sensors) [6–8]. Transparent and photochromic REHO films are produced by either reactive Ar-magnetron sputtering [9] or reactive electron-beam evaporation [10] of a rare-earth metal hydride (REH$_2$) onto a transparent substrate, followed by oxidation in air. The films were reported to crystallize into a face-centered cubic (fcc) structure [11] and it is found that oxygen replaces hydrogen during the oxidation process [12], with strong dependence of the photochromic response on the chemical composition of the film [13]. As oxidation of the thin films of REHO continues in air [14], the chemical composition changes and therefore the optical properties. Upon long time exposure, the films become transparent rare-earth metal oxides hampering applications requiring long time stability.

In-situ composition analysis of photochromic YHO samples (RE = yttrium) under illumination showed that photodarkening can be triggered in high-vacuum conditions (base pressure $\approx 10^{-5}$ Pa) and is not related to significant chemical composition changes ($\geq 1$ at.%) of the films. [15] These results suggest that the photochromic properties are rather linked to e.g. changes in the structure or electronic rearrangements. In our recent work [16], we have observed a columnar-type structure and a coexistence of fcc REH$_2$-like and bixbyite-like RE$_2$(H$_x$O$_y$)$_3$ phases with compressive residual stress on the order of GPa in the films. As it is known that yttrium hydride is photochromic at high pressures [17] we propose a similar mechanism of photodarkening in REHO, accompanied by a photon-induced hydrogen transfer from the bixbyite RE$_2$O$_3$ phase into the fcc REH$_2$ phase. Recent calculations using time-dependent density functional theory based on excited-state molecular dynamic support this hypothesis [18]. In parallel, a possible gas exchange between the film and the environment has been suggested as the key mechanism to the photochromic response [19]. However, a close look at the data presented by the authors [19,20] shows that a change of the relative intensity of the overlapping Bragg peaks of the two phases induced by material transport from one to another can provide an alternative explanation.

In this letter, we show that encapsulated photochromic REHO films with transparent, non-photochromic and thin diffusion-barriers perform equally well as uncapped films. Moreover, we prove that the films sealed against the environment are still photochromic and long-time stable, as oxidation is inhibited by the diffusion barrier. Our results show that material transport *in* and *out* of the film cannot be the reason for the photochromic effect.

## 2. Experimental details

Two batches of two $YH_2$ samples each (total four samples) were reactively grown by Ar-magnetron sputtering onto soda-lime glass substrates (ultrasonic-cleaned microscope slides, 10x10 mm$^2$ and 1 mm thick) using a compact Balzers Union sputtering device. Further details on the synthesis of REHO can be found elsewhere. [21] After the metal-hydride deposition, the samples were exposed to air for oxidation. At this stage, the films change their appearance from opaque to (photochromic) yellowish transparent. After exposure to air, one sample is kept in air, while the other one is capped with either a thin layer of $Al_2O_3$ or $Si_3N_4$ ($\approx$ 20 min of time lag). Both these layers have good antireflective properties, lower absorption of visible light in comparison to YHO. The optical band gap ($E_g$) of amorphous $Al_2O_3$ and $Si_3N_4$ is $E_g \approx$ 5-7 eV [22,23] and $E_g \approx$ 2.7 eV [12] for YHO. The fabrication processes are well established (with low contamination levels) and used in semiconductors and solar cells as diffusion barrier coatings. [24–27]

The $Al_2O_3$ capping layer was deposited in a pure Ar-environment (base pressure $\approx 10^{-3}$ Pa). During deposition, the Ar-pressure was maintained at $\approx$ 5 Pa. The target-substrate distance was 5 cm and the plasma sputtering current 45 mA. Three 200 s sputtering cycles were performed, each one followed by an abrupt oxidation by air. Every sputtering cycle yielded a thin layer of metallic aluminum, and the subsequent oxidation transformed the layer into fully-transparent $Al_2O_3$. For the $Si_3N_4$ capping layer one sample was transferred to a reactive magnetron sputtering system (Von Ardenne CS 730S at MyFab, Uppsala University) equipped with a silicon target (nominal purity > 99.999 %). The Ar:N2 ratio was kept at 20:40 sccm with a 500 W of plasma power (base pressure during grown $\approx$ 0.8 Pa). The resulting deposition rate was $\approx$ 3 Å/sec.

The chemical composition of the samples was obtained by combining results from four ion beam-based methods: Rutherford Backscattering Spectrometry (RBS), Time-of-Flight Elastic Recoil Detection Analysis (ToF-E ERDA), Nuclear Reaction Analysis (NRA), and Elastic Backscattering Spectrometry (EBS). All measurements were carried out at the 5-MV NEC-5SDH-2 tandem accelerator (Tandem laboratory, Uppsala University) [28]. The spectra were analyzed following an iterative approach (see Ref. [29] for further details on the techniques and evaluation procedure), providing composition depth-profiles of the samples, including quantification of trace impurities.

In Figure 1, we show the chemical composition depth-profiles from the ToF-E ERDA analysis. Panels (a) and (b) depict the results from the first batch of twin YHO samples, one being encapsulated with an $Al_2O_3$ layer. Panels (c) and (d) represent depth-profiles from the second batch, i.e., the uncapped and $Si_3N_4$-capped films, respectively. The chemical composition of the YHO of each batch of samples is identical within ± 2 at.% (depth-range to ≈ 480 nm due to limited probing depth of ToF-ERDA). Note that the ToF-E ERDA depth-profiles contain inputs from other methods (see e.g. Ref. [3] for a similar system). Light impurities (C and F) originating from the deposition process are found for all samples (in sum ≤ 5-7 at.%). The encapsulation did not affect the chemical composition of the films and the average [O]/[Y] ratio in the YHO bulk is found to be ≈ 0.51 ($Al_2O_3$ batch) and ≈ 0.43 ($Si_3N_4$ batch), in agreement to the composition range showing photochromism $0.40 \leq [O]/[Y] \leq 1.5$ [12]. From the RBS results (not shown), the total film thicknesses (assuming YHO bulk densities – see supplemental information in Ref. [13]) are found to be 412 nm and 478 nm for the $Al_2O_3$ and $Si_3N_4$ batches, respectively. Moreover, RBS analysis did not reveal any presence of yttrium on the encapsulated sample surface (less than 1 at.%), indicating the absence of holes and cracks in the capping layers. The signal of Y, O and H found in the capping layers in Figure 1 is a consequence of the depth-resolution of the ToF-E ERDA as well as multiple scattering effects (the latter especially for Y). The thicknesses, extracted from the RBS data, of the capping layers are 52 nm and 60 nm for the $Al_2O_3$ and $Si_3N_4$, respectively. Film thicknesses and their chemical composition are summarized in Table 1.

**Table 1.** Summary of the film (and capping) thicknesses and their respective optical properties.

| Diffusion barrier: | $Al_2O_3$ layer | | $Si_3N_4$ layer | |
|---|---|---|---|---|
| | **Uncapped** | **Capped** | **Uncapped** | **Capped** |
| *Cap thickness [nm]* | --- | 52 | --- | 60 |
| *Film thickness [nm]* | 412 | 412 | 478 | 478 |
| *Average [O]/[Y] ratio (bulk)* | 0.51 | 0.50 | 0.45 | 0.44 |
| *Photochromic response [%]* | 18 | 15 | 33 | 36 |
| *Bleaching constant [min]* | 18.5 | 19.3 | 46.1 | 46.4 |

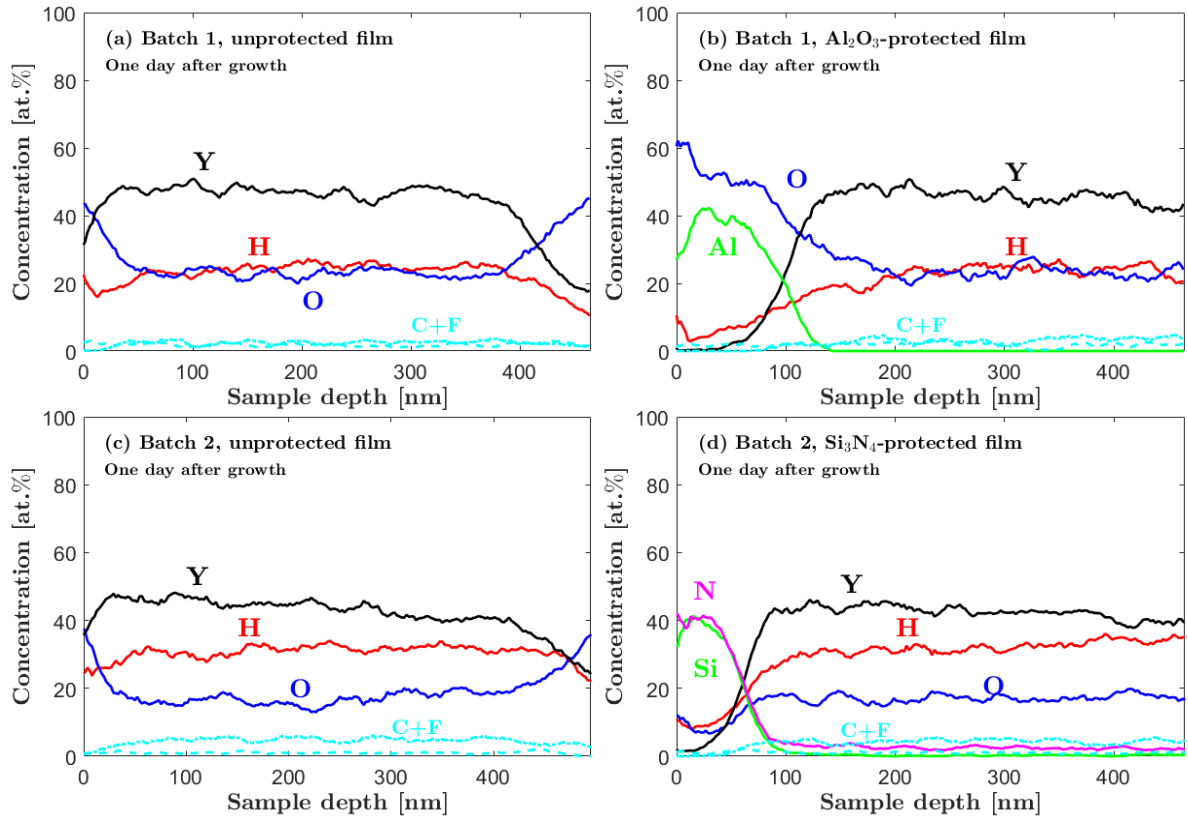

**Figure 1.** Depth-profiles deduced by ToF-E ERDA for YHO samples (uncapped (a) and capped with Al$_2$O$_3$ (b)) from the first batch. Panels (c) and (d) depict data for the second batch of YHO samples (uncapped (c) and capped with Si$_3$N$_4$ (d)).

Optical measurements were carried out using a Perkin Elmer Lambda 35 UV/Vis spectrophotometer equipped with a tungsten halogen and deuterium light source. The optical transmission was calibrated with respect to the transmission of air. Optical scans were done in the wavelength range of [300–1000] nm, at 240 nm/min (slit size of 2 nm). First, the optical transmission is measured while the samples are fully bleached (i.e., initial stage before illumination), followed by a measurement after the samples being photo-darkened by 20 min using a LED array of blue light (wavelength 400 nm and intensity $\approx$ 10 mW/cm$^2$). The photochromic response is defined as the ratio of the averaged optical transmission before and after illumination (integrated in the wavelength range [500 – 900] nm). All samples relaxed to their initial transmission once illumination was stopped. To eliminate the effect of different absorption coefficients in the capping layers, all samples were illuminated through the glass substrate. A potential influence from reflection at the interfaces is negligible.

## 3. Results and discussion

In Figure 2 panels (a) and (b), the optical transmission spectra of the YHO samples before (dashed lines) and after (solid lines) illumination are shown. The measurements were carried out ≈ 6 h after fabrication. For comparison, spectra from pure glass (black solid lines) and the capping layers deposited onto glass (black dashed lines) are shown. The photochromic responses are ≈ 18 % and ≈ 15 %, for the uncovered (blue lines) and $Al_2O_3$-encapsulated YHO (red lines) samples, respectively. For the second batch, a generally higher photochromic response ≈ 33 % and ≈ 36 % was found, for the uncovered (blue lines) and $Si_3N_4$ encapsulated (red lines) YHO samples, respectively. It turns out that encapsulation does not influence the photo-darkening of the films.

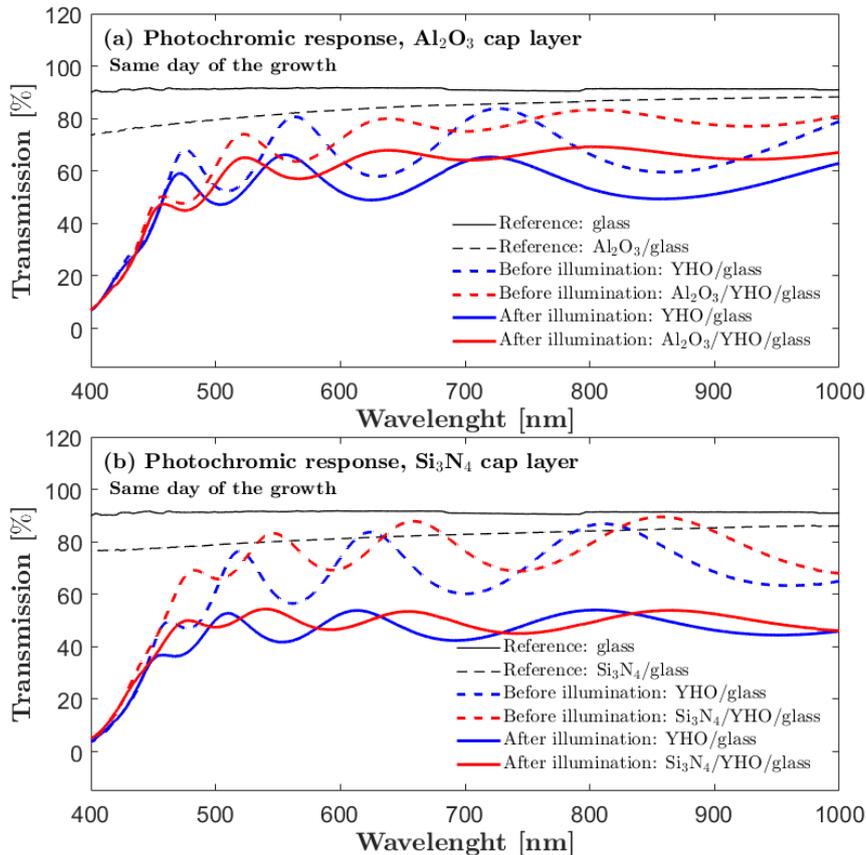

**Figure 2**. Optical transmission as a function of wavelength for the capped and uncapped YHO samples before (dashed line) and after (solid line) illumination ($Al_2O_3$-batch in (a) and $Si_3N_4$-batch in (b)). The black solid and dashed lines represent reference measurements on pure glass and capping layer deposited on glass.

The bleaching process, i.e. i.e. the relaxation from the photo-darkened to the original and transparent stage, showed similar relaxation time constants to the initial transmission (deduced from exponential fits to the time-dependent optical transmission data) of 18.5 min (as-deposited) and 19.3 min (encapsulated) for YHO and from the $Al_2O_3$-batch, and 46.1 min (as-deposited) and 46.4 min (encapsulated) for YHO films from the $Si_3N_4$-batch, respectively.. A summary of the photochromic properties is given in Table 1. As the $Si_3N_4$ capping layer is a stable and oxygen-free diffusion-barrier we can exclude any oxygen uptake or release by REHO films during photodarkening and bleaching.

To investigate the aging and degradation of the films, we have measured the optical response of the samples from the $Al_2O_3$ batch at several instances up to 14 days after the growth. The results are shown in Figure 3, panel (a) for the encapsulated film and panel (b) for unprotected one. The photochromic performance of the protected film remains unchanged, within $\approx$ 2-3 %. For the unprotected YHO film, delamination was observed after a few days of exposure to air, and its optical properties have been degrading even more rapidly. Note, that continuous air exposure of uncapped films may initially result in either an increasing or decreasing photochromic response as this quantity depends on the oxygen concentration and is largest for [O]/[H] $\approx$ 0.8, see Ref. [12]). Our results show that once capped, the oxygen concentration in the YHO film remains constant allowing to tailor design photochromic films with long time stability. Thus, to obtain reproducible and maximized photochromic response optimizing the oxidation process before depositing a capping layer, potentially by controlled oxidation [10], is desirable.

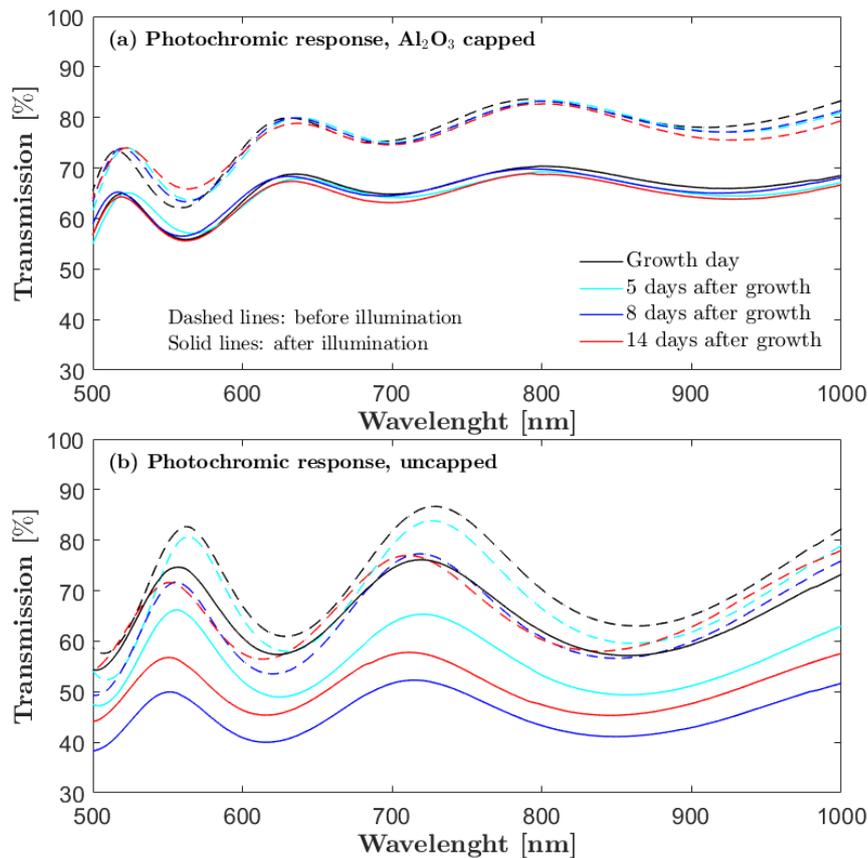

**Figure 3.** Optical transmission spectra showing degradation of the photochromic response for YHO samples (capped with $Al_2O_3$ – panel (a) and as deposited – panel (b)). After encapsulation the films are long time stable.

**4. Summary and conclusions**

We have fabricated photochromic YHO films encapsulated by transparent and non-photochromic diffusion-barrier layers of $Al_2O_3$ and $Si_3N_4$. Composition depth profiling shows that the encapsulation layers are stable, free of pinholes or cracks and only a few tens of nanometers thick. The photochromic properties (photodarkening response and relaxation time) for the encapsulated samples are the same as those of freshly prepared YHO unprotected films but stable over time. From our results, we conclude that material transport to and from the ambient, e.g. release and uptake of oxygen or other species, cannot explain the photochromic effect. In addition, our results demonstrate that photochromic films can be efficiently protected against oxidation and become long time stable, which is a huge step forward towards technological application.

**Acknowledgments**

D.M. acknowledges Visby Programme Scholarships for PhD studies. Support by VR-RFI (#2017-00646_9) and the Swedish Foundation for Strategic Research (SSF, contract RIF14-0053) supporting accelerator operation is gratefully acknowledged.

**Conflict of interest**

The authors declare no financial/commercial conflict of interest.